By Erik Høg, Niels Bohr Institute,
Copenhagen University, ehoeg@hotmail.dk

# The Astrometric Foundation of Astrophysics

The astrometric foundation of astrophysics was revolutionized by the observations 1989-93 with ESA's Hipparcos astrometry satellite. The accuracy of positions, proper motions and distances (parallaxes) for a large number of stars was unprecedented. For distances, e.g., an accuracy of 1.0 per cent or better was obtained for 719 stars. No other star than the Sun had such an accurate distance before Hipparcos. The Tycho-2 Catalogue with positions, proper motions and *B,V* photometry for the 2.5 million brightest stars in the sky has become the preferred catalogue for these stars since its publication in 2000.

The second astrometry satellite, Gaia, was launched by ESA in December 2013 and is scheduled to observe more than one billion stars to a limit of 20 mag during 5-6 years. Distances with 1.0 per cent accuracy will be obtained for 10 million stars; at 14 mag the astrometric accuracy will be about 10 micro-arcseconds. Photometry and radial velocities are also obtained.

The accurate astrometric data expected from Gaia offer the opportunity and the obligation to exploitation by a follow-up all-sky mission. Therefore I submitted a proposal to ESA in May 2013 for a **Gaia successor mission in about twenty years with similar performance as Gaia**. It is important to capitalize on the technical and scientific experience in Europe with Hipparcos and Gaia and on the unique results from these missions. In brief, the two Gaia-like missions would provide an astrometric foundation for all branches of astronomy from the solar system and stellar systems to compact galaxies, quasars and dark matter by data which cannot be surpassed the next 50 years.

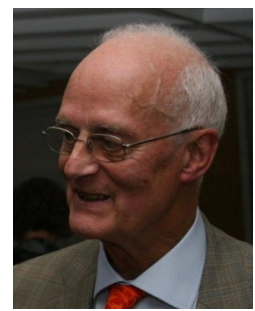

The future of fundamental astrometry must be considered in a time frame of 50 years. A dozen science issues for a Gaia successor mission are presented and in this context the possibilities for absolute astrometry are discussed in a report of 30 pages:

**Absolute astrometry in the next 50 years** at
https://dl.dropbox.com/u/49240691/GaiaRef.pdf

*Absolute astrometry* means data in an inertial coordinate system, the International Celestial Reference System, ICRS.

The three powerful astrometric techniques: VLBI, the MICADO camera on the E-ELT, and the LSST are discussed in the report, documented by literature references and by an extensive correspondence with leading astronomers who readily responded with all the information needed. - *All astronomers are urged to think of their future needs for astrometry and to propose science cases,* in addition to the dozen cases mentioned in the report.

The present author contributed to the Hipparcos/Tycho and Gaia missions during 32 years from 1975. I invented the basic technique of *photon counting astrometry* in 1960 and implemented it on the meridian circle of the Hamburg Observatory. This was the basic detection technique for Hipparcos/Tycho. The basic detection technique in Gaia, however, is direct imaging on CCDs as I proposed for the Roemer astrometry satellite project in 1992.

---

[1] This abstract was originally submitted in Latex for the conference book, but is here reformated to Word because *arXiv* could not accept the resulting PDF.

# Absolute astrometry in the next 50 years

Erik Høg - ehoeg@hotmail.dk

Niels Bohr Institute, Copenhagen, Denmark

**Abstract:** With the Gaia astrometry satellite in orbit since December 2013 it is time to look at the future of fundamental astrometry - a time frame of 50 years is needed in this matter. A dozen science issues for a Gaia successor mission in twenty years are presented and in this context the possibilities for absolute astrometry with milli-arcsec or sub-mas accuracies are discussed. In brief, the two Gaia-like missions would provide an astrometric foundation for all branches of astronomy from the solar system and stellar systems, including exo-planet systems, to compact galaxies, quasars and dark matter by data which cannot be surpassed the next 50 years. The presentation is based especially on the paper: "Absolute astrometry in the next 50 years" at http://arxiv.org/abs/1408.2190 and on a discussion herein of the detection of Exo-Jupiters and Saturns from two Gaia-like missions.

Firt presented as invited review in July 2014 at the 75 year anniversary of the Kiev Main Astronomical Observatory and in September at the annual meeting of the Astronomische Gesellschaft in Bamberg.

---

**The astrometric foundation of astrophysics**

**Top science**

*from a new Gaia-like mission vs. a single Gaia:*

- *Imaging of radio/optical sources etc.:*
  *Positions 50 years from now* >20 times smaller errors
- *Dynamics of Dark Matter etc. from stellar proper motions:*
  *Tangential velocities with 10 times smaller errors*
  *in 30 times larger volume*
- *Stellar distances in >3 times larger volume*
- *Exoplanets: Periods up to 40 years, vs. Gaia P<10 yrs*
- *Quasars solely by zero motions: 100 times cleaner sample*
- *Solar system: orbits, asteroid masses…*
- *Astrometry and photometry with 0."1 resolution*
- *Astrometric binaries. Common proper motion pairs. Etc. etc.*

---

**Figur 1.** A long-lasting astrometric foundation of astrophysics will be obtained by a new Gaia-like mission launched 20 years after the first. For example, in 2066, 50 years from now, positions predicted from two missions will have 20 times smaller errors than from Gaia alone.